\documentclass[twocolumn,twoside,amsmath,amssymb,superscriptaddress,showpacs,nofootinbib,aps]{revtex4-1}
       
\usepackage{slashed}
\usepackage{amsmath}
\usepackage{amsthm}
\usepackage{mathtools}
\usepackage[utf8]{inputenc}
\usepackage[T1]{fontenc}
\usepackage{graphicx}
\usepackage{epsfig}
\usepackage{amssymb}
\usepackage{dcolumn}
\usepackage{bm}
\usepackage{cancel}
\usepackage{color}
\usepackage{cancel}
\usepackage{multirow}
\usepackage{hyperref}
\usepackage{cleveref}

\newcommand{\ba}{\begin{array}}
\newcommand{\ea}{\end{array}}
\def\br{\begin{eqnarray}}
\def\er{\end{eqnarray}}
\def\be{\begin{equation}}
\def\ee{\end{equation}}

\def\({\left(}
\def\){\right)}

\def\<{\left\langle}
\def\>{\right\rangle}

\begin{document}


\title{Nonperturbative gluon exchange in $pp$ elastic scattering at TeV energies}

\author{G.~B.~Bopsin}
\email{gustavo.bopsin@ufrgs.br}

\affiliation{Instituto de F\'isica, Universidade Federal do Rio Grande do Sul, Caixa Postal 15051, 91501-970, Porto Alegre, RS, Brazil}

\author{E.~G.~S.~Luna}
\email{luna@if.ufrgs.br}

\affiliation{Instituto de F\'isica, Universidade Federal do Rio Grande do Sul, Caixa Postal 15051, 91501-970, Porto Alegre, RS, Brazil}

\author{A.~A.~Natale} 
\email{adriano.natale@unesp.br}

\affiliation{Instituto de F{\'i}sica Te\'orica - UNESP, Rua Dr. Bento T. Ferraz, 271,\\ Bloco II, 01140-070, S\~ao Paulo, SP, Brazil}

\author{M. Pel\'aez} 
\email{mpelaez@fing.edu.uy}

\affiliation{Instituto de F\'{\i}sica, Facultad de Ingenier\'{\i}a, Universidad de la Rep\'ublica, \\
J.H. y Reissig 565, 11000 Montevideo, Uruguay}

\begin{abstract}
We investigate the two-gluon-exchange model of the Pomeron using nonperturbative gluon propagators characterized by a dynamical mass scale. We present the results for an analysis of the available $pp$ differential cross section data at TeV energies which accounts for dynamical gluon masses obtained from a non-linear version of the Schwinger-Dyson equations. We show that our two-gluon exchange model gives an excellent description of the LHC data, provided we demand the Reggeization of the scattering amplitude and make a suitable choice for the convolution of proton wave functions.

\end{abstract}



\maketitle

\section{Introduction}

It remains a challenge for elementary particle physics to understand the QCD nature of the Pomeron, a colorless state having the quantum numbers of the vacuum. It has been known for a long time that the behavior of the hadronic cross sections at high energies in the soft regime is well described in the framework of Regge theory, in which the behavior of the scattering amplitude is driven by singularities of the amplitude in the complex plane of angular momentum $j$. In the simplest scenario the scattering amplitude is dominated by an isolated pole at $j=\alpha_{\Bbb P}(t)$, resulting in an amplitude ${\cal A}(s, t) \propto s^{\alpha_{\Bbb P} (t)}$, where $\alpha_{\Bbb P}(t)$ is the Pomeron pole trajectory. The ultimate goal is to incorporate QCD concepts into the Pomeron construction in order to reproduce at least some of the phenomenological features of the soft Pomeron.

Various attempts using QCD ideas have been made to study the soft Pomeron and, 
since the work of Low and Nussinov \cite{low001,nussinov001}, it has been realized that the lowest-order QCD construction possessing the correct Pomeron quantum numbers ($C=+1$, color singlet) is the two-gluon exchange. The first perturbative calculations using such a model, although not successful in describing the scattering data available at the time, were instructive in highlighting some phenomenological possibilities \cite{gunion001,ryskin001,richards001}. In these calculations, the scattering amplitude was written as
\begin{eqnarray}
{\cal A}(s,t)=is\frac{8}{9}n_{p}^{2}\alpha_{s}^{2}\left[ T_{1} - T_{2} \right] ,
\label{amplit001}
\end{eqnarray}
where $T_{1}$ ($T_{2}$) represent the contribution when both gluons attach to the same quark (to different quarks) within the proton. Here $n_{p}=3$ is the number of quarks in the proton, and $\alpha_{s}$ is the canonical strong coupling. Among the main results of these calculations, we have a total cross section that is constant in $s$ \cite{gunion001,ryskin001} as well as an amplitude that decreases much more rapidly with increasing $|t|$ than that generated by single-Pomeron exchange \cite{richards001}. Most importantly, the perturbative calculation of the elastic hadron-hadron scattering amplitude through a two-gluon exchange is invariably accompanied by a singularity at $-t=0$. Since the origin of this singularity is the pole in the gluon propagator at $q^{2}=0$, Landshoff and Nachtmann (LN) suggested that the gluon propagator is intrinsically modified in the infrared region \cite{LN001}. They noticed that the singularity present in the two-gluon exchange calculation of the hadron-hadron scattering is eliminated if the gluon propagator is finite at $q^{2}=0$. In the LN model, the Pomeron exchange corresponds to the two-gluon exchange. These two gluons couple predominantly to the same quark in the hadron, and this exchange behaves like a $C=+1$ photon-exchange diagram with an amplitude
\begin{eqnarray}
i\beta_{0}^{2}\left( \bar{u} \gamma_{\mu} u\right) \left( \bar{u} \gamma^{\mu} u\right),
\end{eqnarray}
where $\beta_{0}$ represents the strength of the Pomeron coupling to quarks, being given by
\begin{eqnarray}
\beta_{0}^{2} = \frac{1}{36 \pi^{2}} \int d^{2}k \left[ g^{2} D(k^{2})  \right]^{2} .
\label{eqln002}
\end{eqnarray}
It is worth remarking on the fact that the convergence of the integral in Eq. (\ref{eqln002}) requires a {\it nonperturbative} gluon propagator, i.e. a propagator in which the infrared pole at $q^{2}=0$ is removed by some nonperturbative mechanism. Very soon after the introduction of these ideas, several phenomenological consequences have been discussed in the literature \cite{landshoff001,ross001,natale002}. For example, using nonperturbative gluon propagators in LN-type models, it was possible to describe low-energy data on $J/\Psi-$nucleon total cross section, to compute an estimate for the differential cross section of the process $\gamma\gamma \to J/\Psi J/\Psi$, and to compute the elastic differential cross section for $pp$ scattering at $\sqrt{s}=53$ GeV. 

After precise measurements of elastic $pp$ scattering at LHC have been released, an LN-inspired approach based on the refined Gribov-Zwanziger framework and massive Cornwall-type gluon propagator was used in the calculation of the differential cross section at $\sqrt{s}=$ 7, 8, and 13 TeV \cite{dudal001}. Surprisingly, the calculation is in complete disagreement with the experimental data, providing a reasonable description of $d\sigma /dt$ again only at low energies, namely $\sqrt{s} =$ 53 TeV. It is important to be absolutely clear that the contribution of the Pomeron component to $\chi^{2}$ is completely dominant in the LHC regime \cite{luna002,luna004}. In other words, at TeV energies the Reggeon (non-Pomeron) contributions are negligible, and it seems very plausible that any Pomeron-type model should therefore work precisely at the LHC energies. Hence there is every reason to believe that the LHC energy regime sets up the stage for carrying out a systematic study of the LN Pomeron.

In this Letter we show that an LN-type model can, in fact, describe the LHC data with great accuracy, provided we make an appropriate choice for the convolution of proton wave functions and demand the Reggeization of the scattering amplitude.

\section{The model}

One of the remarkable features of non-Abelian gauge theories is the Reggeization of elementary particles \cite{grisaru001,lipatov001,fadin001}, particularly in the case of QCD. Gluon Reggeization turns out to be of central importance at high energies since only cross sections for processes involving the exchange of gluons in the $t$-channel do not fade away as $s$ increases; in each fixed order of perturbation Reggeized gluons completely dominate the amplitudes for such processes. Furthermore, the gluon Reggeization plays a central role in the derivation of the BFKL equation \cite{bfkl}. This equation describes the leading logarithmic evolution of gluon ladders in $\ln s$, in which the vertical lines are Reggeized gluons. This means that these gluonic lines are not composed of bare gluons whose propagators (in the Feynman gauge) are given by
\begin{eqnarray}
D_{\mu\nu}(q^{2}) = -i \frac{g_{\mu\nu}}{q^{2}},
\end{eqnarray}
but rather composed of gluons whose propagator is
\begin{eqnarray}
D_{\mu\nu}(\hat{s},q^{2}) = -i \frac{g_{\mu\nu}}{q^{2}} \left( \frac{\hat{s}}{{\bf k}^{2}}   \right)^{\epsilon_{G}(q^{2})},
\end{eqnarray}
where ${\bf k}^{2}$ is a typical transverse momentum, $\hat{s}$ is the square of the total center-of-mass of the particles which exchange the Reggeized gluon, and $\alpha_{G}(q^{2})=1+\epsilon_{G}(q^{2})$ is the Regge trajectory of the gluon. Thus in the case of color-octet exchange, in the limit $s \gg |t|$, the BFKL equation exhibits a pole solution, corresponding to a single Reggeized gluon propagating in the $t$-channel. Similarly, in the case of a color-singlet exchange, a gluon ladder configuration corresponds to a bound state of gluons, namely the BFKL Pomeron.

More generally, if the amplitude ${\cal A}(s,t)$ for a process involving the exchange in the $t$-channel of the quantum numbers of a particle of mass $M$ and spin $j$ behaves asymptotically as ${\cal A}(s,t) \propto s^{\alpha (t)}$, it is said that we are treating with a `Reggeized' particle, where $\alpha (t)$ is the trajectory of the particle; in particular, the particle lies on the trajectory, i.e. $\alpha (M^2) = j$. Following this line of thought, one might then be led to consider changes of the form $s \to s^{\alpha (t)}$ as a phenomenological procedure for the Reggeization of scattering amplitudes. In our case, a simple change $s \to s^{\alpha_{\Bbb P} (t)}$ in the amplitude (\ref{amplit001}) would, on this analogy, lead us to expect a Reggeized version of the LN amplitude. Thus, by considering the LN-Pomeron Reggeization, one verifies that the scattering amplitude (\ref{amplit001}) may be rewritten as  
\begin{eqnarray}
 {\cal A}(s,t) = is^{\alpha_{\Bbb P}(t)} \frac{1}{\tilde{s}_{0}}\frac{8}{9}n_{p}^{2} [ \tilde{T}_{1} - \tilde{T}_{2} ] ,
\label{ampliyyy}
\end{eqnarray}
with
\begin{widetext}
\begin{eqnarray}
\tilde{T}_{1} = \int_{0}^{s} d^{2}k \, \bar{\alpha}\left( \frac{q}{2} + k \right) D\left( \frac{q}{2} + k \right) \bar{\alpha}\left( \frac{q}{2} - k \right) D\left( \frac{q}{2} - k \right) \left[ G_{p}(q,0)  \right]^{2} ,
\label{t1}
\end{eqnarray}
\begin{eqnarray}
\tilde{T}_{2} = \int_{0}^{s} d^{2}k \, \bar{\alpha}\left( \frac{q}{2} + k \right) D\left( \frac{q}{2} + k \right) \bar{\alpha}\left( \frac{q}{2} - k \right) D\left( \frac{q}{2} - k \right) G_{p}\left( q,k - \frac{q}{2}\right) \left[ 2 G_{p}(q,0) - G_{p}\left( q,k - \frac{q}{2}\right)  \right] .
\label{t2}
\end{eqnarray}
\end{widetext}
Here $\alpha_{\Bbb P}(t)=1+\epsilon + \alpha^{\prime}_{\Bbb P}t$ is the LN-Pomeron trajectory, $\tilde{s}_{0} \equiv s_{0}^{\alpha_{\Bbb P}(t)-1}$ (where the mass scale $s_{0} \equiv 1$ GeV$^{2}$ have been introduced to get the dimension of the total cross section, $\sigma_{tot}(s)$, right), and $G_{p}(q,k)$ is a convolution of proton wave functions,
\begin{eqnarray}
G_{p}\left( q,k \right) = \int d^{2}p \, d\alpha \, \psi^{*}(\alpha, p)\, \psi (\alpha, p - k - \alpha q) ,
\end{eqnarray}
where the wave function $\psi (\alpha, p)$ is the amplitude for the quark to have transverse momentum $p$ and fraction $\alpha$ of the longitudinal momentum. In this picture $G_{p}(q,0)$ is simply the proton elastic form factor, $F_{1}(q^{2})$. We estimate $G_{p}\left( q,k - \frac{q}{2}\right)$ assuming a proton wave function peaked at $\alpha=1/3$ and using \cite{ross001}
\begin{eqnarray}
G_{p}\left( q,k - \frac{q}{2} \right) = F_{1}\left( q^{2} +9 \left| k^{2} -\frac{q^{2}}{4} \right|  \right).
\end{eqnarray}

The expressions for $\tilde{T}_{1}$ and $\tilde{T}_{2}$ include the nonperturbative QCD information. The nature of the coupling $\bar{\alpha}(q^{2})$ and the gluon propagator $D(q^{2})$ will be discussed in the next section. Notice that, in contrast to (\ref{amplit001}), in the expression (\ref{ampliyyy}) we have inserted the couplings into the integrals (\ref{t1})  and (\ref{t2}). In this form, it is particularly evident that we are using the prescribed calculational scheme, as dictated by the Eq. (\ref{eqln002}): the strength of the Pomeron depends on the product of the coupling $g^{2}(k^{2})$ with the propagator $D(k^{2})$. Furthermore, it is the same procedure used in lattice QCD calculations, where the Pomeron's strength is proportional to the integral $\int d^{2}p\, [g^{2}_{eff}(p^{2})D_{lat}(p^{2})]^{2}$.

The total cross section $\sigma_{tot}(s)$ and the elastic differential cross section $d\sigma/dt$ are, in terms of the amplitude (\ref{ampliyyy}), given by
\begin{eqnarray}
\sigma_{tot}(s) = \frac{\textnormal{Im}\,{\cal A}(s, t=0)}{s}  ,
\end{eqnarray}
\begin{eqnarray}
\frac{d\sigma}{dt}(s, t) = \frac{\left| {\cal A}(s,t) \right|^{2}}{16 \pi s^{2}}  .
\end{eqnarray}

\section{The nonperturbative input}

It is a currently accepted scenario that the nonperturbative dynamics of QCD may generate a dynamical mass $m(q^2)$ for the gluons \cite{aguilar001}. Large-volume lattice QCD calculations indicate that such an effective momentum-dependent mass does arise in both SU(2) \cite{lattice002} and SU(3) \cite{lattice001} simulations. The lattice calculations also reveal a finite gluon propagator in the infrared region \cite{othergauge001}. Moreover, according to the Schwinger-Dyson equations, which in the continuum govern the nonperturbative dynamics of the gluon propagator, a finite gluon propagator corresponds to a dynamically massive gluon \cite{smekal001}.

The phenomenon of dynamical gluon mass generation is intimately related to the concept of QCD effective charge \cite{cornwall001,aguilar002,quinteros001}. A QCD effective charge $\bar{\alpha}(q^{2})$ is a nonperturbative generalization of the perturbative running coupling $\alpha_{s}(q^{2})$ and can be obtained, for example, within the framework of pinch technique \cite{cornwall001,cornwall002,cornwall003}: 
the Schwinger-Dyson solutions for the gluon self-energy $\hat{\Delta}(q^{2})$ (in the background-field method \cite{abbott001}) are
used to form a renormalization-group invariant quantity defined by
\begin{eqnarray}
\hat{d}(q^{2})=g^{2}\hat{\Delta}(q^{2}),
\label{eqan03}
\end{eqnarray}
where $g$ is the gauge coupling. From this quantity, the effective charge may then be defined as
\begin{eqnarray}
\bar{\alpha}(q^{2}) = \left[ q^{2} + m^{2}(q^{2})  \right] \hat{d}(q^{2}),
\label{eqan02}
\end{eqnarray}
where $m(q^{2})$ is the gluon dynamical mass. The inverse of $\hat{d}(q^{2})$ may be written as
\begin{eqnarray}
\hat{d}^{-1}(q^{2}) = \frac{\left[ q^{2} + m^{2}(q^{2})  \right]}{\bar{\alpha}(q^{2})},
\label{eqan04}
\end{eqnarray}
where now
\begin{eqnarray}
\frac{1}{\bar{\alpha}(q^{2})} = b_{0} \ln \left( \frac{q^{2} + m^{2}(q^{2})}{\Lambda^{2}}  \right),
\label{eqan05}
\end{eqnarray}
where $b_{0}=\beta_{0}/4\pi = (33-2n_{f})/12\pi$ is simply the first coefficient of the QCD $\beta$ function (here $n_{f}$ is the number of flavors) and $\Lambda$ is the dimensionful QCD parameter. Note that if $q^{2} + m^{2}(q^{2}) \to p^{2}$ in the argument of the logarithm of (\ref{eqan05}), we obtain the expression for the leading order (LO) perturbative QCD coupling, namely
\begin{eqnarray}
\frac{1}{\alpha_{s}^{LO}(p^{2})} = b_{0} \ln \left( \frac{p^{2}}{\Lambda^{2}}  \right);
\label{eqan06}
\end{eqnarray}
thus, in practice, the QCD
effective charge can be directly obtained by saturating the LO perturbative strong coupling $\alpha_{s}^{LO}(q^{2})$, namely
\begin{eqnarray}
\bar{\alpha}(q^{2}) = \left. \alpha_{s}^{LO}(q^{2}) \right|_{q^{2} \to q^{2} + m^{2}(q^{2})} .
\label{eqan07}
\end{eqnarray}
If the Schwinger-Dyson equations preserve the multiplicative renormalizability, the same procedure can be used to build a next-to-leading order effective charge \cite{luna003}.

Functional forms of the gluon dynamical mass $m(q^{2})$ and of the nonperturbative gluon propagator $D_{\mu\nu}$ were found by Cornwall using the pinch technique in order to derive a gauge invariant Schwinger-Dyson equation for the triple gluon vertex and gluon propagator \cite{cornwall001}. Specifically, the gluon propagator $D_{\mu\nu} = -ig_{\mu\nu}D(q^{2})$ obtained from a gauge-invariant set of diagrams for the Schwinger-Dyson has the scalar factor given by 
\begin{eqnarray}
D^{-1}(q^2) = \left[ q^2 + m^2(q^2) \right]b g^{2} \ln \left[ \frac{q^2 + 4m^2(q^2)}{\Lambda^2} \right]\! 
\label{propgt002}
\end{eqnarray}
in Euclidean space, with the dynamical gluon mass given by
\begin{eqnarray}
m^{2}(q^{2}) = m_{g}^{2} \left[ \frac{\ln \left( \frac{q^{2}+4 m_{g}^{2}}{\Lambda^{2}} \right)}{
\ln \left( \frac{4 m_{g}^{2}}{\Lambda^{2}} \right)} \right]^{-12/11} ,
\label{eqlog}
\end{eqnarray}
where $b=b_{0}/4\pi$ and $m_{g}^{2}=m^{2}(0)$. The Cornwall expression (\ref{eqlog}) is a special case of a logarithmic running mass $m^{2}_{log}(q^{2})$, found in a more recent study using a non-linear version of the Schwinger-Dyson equation for the gluon self-energy \cite{agpapa}, given by
\begin{eqnarray}
m^{2}_{log}(q^{2}) = m_{g}^{2} \left[ \frac{\ln \left( \frac{q^{2}+\rho m_{g}^{2}}{\Lambda^{2}} \right)}{
\ln \left( \frac{\rho m_{g}^{2}}{\Lambda^{2}} \right)} \right]^{-1-\gamma_{1}} ,
\label{eqlog2}
\end{eqnarray}
where $\gamma_{1} = -6(1+c_{2}-c_{1})/5$; here $c_{1}$ and $c_{2}$ are parameters related to the ansatz for the fully dressed three-gluon vertex employed in
numerical analyses of the gluon self-energy. Their values are constrained by
a ``mass condition'', which controls the behavior of $m^{2}_{log}(q^{2})$ in the ultraviolet region, namely $c_{1} \in [0.15,0.4]$ and $c_{2} \in [-1.07,-0.92]$.  The parameters $m_{g}$ and $\rho$, which control the behavior of the dynamical mass in the infrared region, are also
constrained by the mass condition to lie in the intervals $m_{g} \in [300, 800]$ and
$\rho \in [1.0, 8.0]$ MeV \cite{agpapa}. 

Another possible asymptotic behavior for the dynamical gluon mass, also obtained at the level of a non-linear Schwinger-Dyson equation, is given by the power-law running mass
\begin{eqnarray}
m^{2}_{pl}(q^{2}) = \frac{m_{g}^{4}}{q^{2}+m_{g}^{2}} \left[ \frac{\ln \left( \frac{q^{2}+  \rho m_{g}^{2}}{\Lambda^{2}} \right)}{
\ln \left( \frac{\rho m_{g}^{2}}{\Lambda^{2}} \right)} \right]^{\gamma_{2}-1}  \, ,
\label{eqpowerlaw}
\end{eqnarray}
where $\gamma_{2} = (4 + 6 c_{1})/5$, with the same type of mass condition now imposing $c_{1} \in [0.7,1.3]$. Here the $\rho$ and
$m_{g}$ parameters are constrained to lie in the same interval as the logarithmic case, namely $\rho \in [1.0, 8.0]$ and $m_{g} \in [300, 800]$ MeV \cite{agpapa}.
We fix $\rho = 4$, $\gamma_{1}=0.084$, and $\gamma_{2}=2.36$ in our analyses since these values are the ones that give the smallest value of $\chi^{2}/\nu$, where $\nu$ is the number of degrees of freedom (DoF).


Given the running behavior of the dynamical gluon masses, $m^{2}_{log}(q^{2})$ and $m^{2}_{pl}(q^{2})$, the QCD effective charge $\bar{\alpha}_{i}(q^{2})$ is written as
\begin{eqnarray}
\bar{\alpha}_{i}(q^{2}) = \frac{1}{b_{0}\ln\left(\frac{q^{2} +
4m^{2}_{i}(q^{2})}{\Lambda^{2}}\right)} ,
\label{ansatz3}
\end{eqnarray}
where $i = log$, $pl$. Finally, combining all these results, we found an expression for $\bar{\alpha}_{i}(q^2)D(q^{2})$ that guarantees the convergence of the integrals (\ref{t1}) and (\ref{t2}), namely
\begin{eqnarray}
\frac{1}{\bar{\alpha}_{i}(q^2)D(q^2)} = b_{0}\left[ q^2 + m^2_{i}(q^2) \right] \ln \left[ \frac{q^2 + 4m^2_{i}(q^2)}{\Lambda^2} \right]\! , \nonumber \\
\end{eqnarray}
where we have used $g^{2}=4\pi \bar{\alpha}_{i}(q^2)$ in the expression (\ref{propgt002}). One very important point to note is that $\bar{\alpha}_{log}(q^{2})$ and $\bar{\alpha}_{pl}(q^{2})$ tame the Landau pole, i.e. they exhibit infrared fixed points as $q^{2} \to 0$. In a mathematical sense, these QCD effective charges belong to the same class of holomorphic couplings \cite{cvetic001}.

\section{results and discussion}

The LHC has performed very precise measurements of diffractive processes that provide a unique constraint on the behavior of the scattering amplitude at high energies. These measurements (and more especially total and differential cross sections from ATLAS and TOTEM experiments) have an accuracy sensitive to nonperturbative physics, allowing us to study the LN Pomeron in more detail.
However, these experimental results reveal some tension between the TOTEM and ATLAS measurements. For example, if we compare the TOTEM result for $\sigma_{tot}^{pp}$ at $\sqrt{s}=7$ TeV, $\sigma_{tot}^{pp}=98.58\pm 2.23$ \cite{antchev001}, with the most precise value measured by ATLAS at the same
energy, $\sigma_{tot}^{pp}=95.35\pm 1.36$ \cite{atlas001}, the difference between the values, assuming that the
uncertainties are uncorrelated, corresponds to 1.4 $\sigma$; if we compare the ATLAS result for the total cross section at $\sqrt{s}=8$ TeV, $\sigma_{tot}^{pp}=96.07\pm 0.92$ \cite{atlas002}, with the lowest value measured by TOTEM at the same center-of-mass energy, $\sigma_{tot}^{pp}=101.5\pm 2.1$ \cite{antchev002}, we see an even more significant difference: 2.6 $\sigma$.
This strong disagreement clearly indicates the possibility of different scenarios for the rise of the total cross section and, consequently, for the parameters of the LN Pomeron.

Thus, in order to investigate
the tension between the TOTEM and ATLAS results in a quantitative way, we carry out global fits to $pp$ differential cross section data considering two distinct ensembles of data with either the TOTEM or the ATLAS measurements. This ``ensemble-selection'' approach is statistically well-founded and
has been used for the first time in the study of cosmic-ray data discrepancies and their effects on the predictions of $pp$ total cross sections at high energies \cite{luna011}. The procedure was later used in the study of Tevatron tension between the CDF and E710/E811 data and its effect on extrema bounds of the soft Pomeron intercept \cite{luna012}. As a result, in all the cases, a very clear distinction among asymptotic values of $\sigma_{tot}^{pp}$ has emerged. As we will see, the discrepancies between the TOTEM and ATLAS data result in distinct values for the LN Pomeron parameters, which in turn also lead to different asymptotic scenarios for $\sigma_{tot}^{pp}$. It follows that the two LHC ensembles for data reductions can be defined and denoted as

{\bf Ensemble A}: ATLAS data on $\frac{d\sigma}{dt}$ at 7, 8, and 13 TeV;

{\bf Ensemble T}: TOTEM data on $\frac{d\sigma}{dt}$ at 7, 8, and 13 TeV.

Once we have defined our data sets, we turn to the phenomenology and carry out global fits to the Ensemble A \cite{atlas001,atlas002,atlas003} and to the Ensemble T \cite{TOTEM001,TOTEM005,TOTEM010} with $|t|_{min} \leq |t|\leq 0.2$ GeV$^{2}$, where the statistic and systematic errors of the data are added in quadrature. We have adopted $|t|_{min} \sim 10 |t|_{int}$, where $|t|_{int} = 0.071/\sigma_{tot}$, since in this region the nuclear scattering dominates \cite{luna004}. The choice for the upper limit on $|t|$ interval will be made clearer in the discussion of the convolution of proton wave functions which follows.
In all the fits to the experimental data we use a $\chi^{2}$ fitting procedure, where the value of $\chi^{2}_{min}$ is distributed as a $\chi^{2}$ distribution with $\nu$ degrees of freedom. The fits are performed adopting an interval $\chi^{2}-\chi^{2}_{min}$ corresponding to 90\% confidence level (CL).

As indicated in Section I, a good description of the differential cross section at TeV energies requires, besides the Reggeization of the scattering amplitude, a more sophisticated version of the convolution of proton wave functions. This is necessary in order to take account of the fact that the $d\sigma /dt$ data at LHC show a significant deviation from an exponential in the small $|t|$ region, as first observed by the TOTEM Collaboration \cite{TOTEM005,TOTEM003,TOTEM009}. As a result, the value found for the nuclear slope $B$ (using an exponential fit at low $|t|$) can be considered as an average $B$, since the high value for $\chi^{2}/DoF$ in the TOTEM fit shows the exponential model as an oversimplified description of the data \cite{TOTEM009}. To obtain a better fit, the TOTEM Collaboration has generalized the pure exponential to a cumulant expansion,
\begin{eqnarray}
\frac{d\sigma}{dt}(t) = \left. \frac{d\sigma}{dt} \right|_{t=0} \exp \left( \sum_{n=1}^{N_{b}} b_{n} t^{n}  \right) .
\end{eqnarray}
Here the $N_{b}=1$ case corresponds to the pure exponential. A satisfactory description of the data at $\sqrt{s}=13$ TeV was achieved in the case $N_{b}=3$, with $\chi^{2}/DoF=1.22$ and $p-\textnormal{value} = 8.0$ \% , using data with $|t|_{max} = 0.15$ GeV$^{2}$, which corresponds to the largest interval before $d\sigma/dt$ accelerates its decrease towards the dip region \cite{TOTEM010,TOTEM009}. 
From considerations based on this observed low-$|t|$ behavior of $d\sigma/dt$ at arbitrarily high energies, we propose the following convolution of proton wave functions at $k^{2}=0$ (i.e. the form factor):
\begin{eqnarray}
G_{p}(q, 0) = F_{1}(q^{2}) = \exp \left[ - \left( \sum_{n=1}^{N_{a}} a_{n} |t|^{n}  \right) \right],
\label{formfac01}
\end{eqnarray}
where $-t=q^{2}$. We investigate three cases for the cumulant expansion (\ref{formfac01}), namely $N_{a}=$ 1, 2, and 3. Our philosophy is to adopt the standard statistical $\chi^{2}$ test in order to evaluate the relativity plausibility of these cases in light of LHC data. More specifically, we consider different cumulant cases and the effectiveness of these choices in describing the $d\sigma /dt$ data sets.
Since the TOTEM cumulant analysis of the nuclear slope has been performed using elastic differential cross section data in the interval $0 \leq |t| \leq 0.15$ GeV$^{2}$, in our analyses, we fit to the $d\sigma^{pp}/dt$ data with $|t|\leq 0.2$ GeV$^{2}$, i.e. we place our upper limit on the $|t|$ interval in a value close to the one adopted by TOTEM.

We have first observed that the fit in the case $N_{a}=1$ is not supported by either of the two ensembles of data. However, the $N_{a}=2$ case provides a very good description of the $d\sigma /dt$ data for both ensembles. Following the philosophy of using the minimum number of free parameters, our model, therefore, adopts the case $N_{a}=2$ for the cumulant expansion. This means that the model has 4 free parameters: $m_{g}$, $\epsilon$, $a_{1}$, and $a_{2}$. In this case the interval corresponding to 90\% CL is simply $\chi^{2}-\chi^{2}_{min}=7.78$. Regarding the other parameters of the model, the slope of the LN Pomeron trajectory, $\alpha^{\prime}_{{\Bbb P}}$, is fixed at the value 0.25 GeV$^{-2}$; this value is in agreement with that usually obtained for the soft Pomeron in Regge-model analyses. Furthermore, in all the fits, we fix $n_{f}=3$ and $\Lambda=$ 284 MeV, since these values are the same ones adopted in other calculations of strongly interacting processes \cite{luna003,luna005}: our purpose is to keep, whenever possible, these two parameters fixed at the same values adopted in other phenomenological analyses in order to focus exclusively on the behavior of the dynamical gluon mass $m_{g}$ and, in this way, to verify if there is any universality in its value.

The values of the parameters of the LN Pomeron in the case of logarithmic (power-law) dynamical mass, determined by fits to Ensemble A and Ensemble T, are listed in Table I (Table II). The curves of the differential cross sections, compared with the experimental data, are shown in Figures 1 (Ensemble A) and 2 (Ensemble T). In these Figures, the solid and dashed curves are the results obtained using $m_{log}(q^{2})$ and $m_{pl}(q^{2})$, respectively.

\begin{table}
\centering
\caption{The values of the LN Pomeron obtained in fits to $d\sigma^{pp}/dt$ data using the logarithmic dynamical mass $m_{log}(q^{2})$ (see Eq. (\ref{eqlog2})).}
\begin{ruledtabular}
\begin{tabular}{cccccc}
 & Ensemble A & Ensemble T \\
\hline
$m_{g}$ (GeV) & $0.356\pm$0.025 & 0.380$\pm$0.023 \\
$\epsilon$ & 0.0753$\pm$0.0024 & 0.0892$\pm$0.0027 \\
$a_{1}$ (GeV$^{-2}$) & 1.373$\pm$0.017 & 1.491$\pm$0.019 \\
$a_{2}$ (GeV$^{-4}$) & 2.50$\pm$0.53 & 2.77$\pm$0.60 \\
\hline
$\nu$ & 108 & 328 \\
$\chi^{2}/\nu$ & 0.71 & 0.67 \\
\end{tabular}
\end{ruledtabular}
\label{tab002}
\end{table}

\begin{table}
\centering
\caption{The values of the LN Pomeron obtained in fits to $d\sigma^{pp}/dt$ data using the power-law dynamical mass $m_{pl}(q^{2})$ (see Eq. (\ref{eqpowerlaw})).}
\begin{ruledtabular}
\begin{tabular}{cccccc}
 & Ensemble A & Ensemble T \\
\hline
$m_{g}$ (GeV) & 0.421$\pm$0.030 & 0.447$\pm$0.026 \\
$\epsilon$ & 0.0753$\pm$0.0025 & 0.0892$\pm$0.0027 \\
$a_{1}$ (GeV$^{-2}$) & 1.517$\pm$0.019 & 1.689$\pm$0.021 \\
$a_{2}$ (GeV$^{-4}$) & 2.05$\pm$0.45 & 1.70$\pm$0.51 \\
\hline
$\nu$ & 108 & 328 \\
$\chi^{2}/\nu$ & 0.64 & 0.90 \\
\end{tabular}
\end{ruledtabular}
\label{tab002}
\end{table}

\begin{figure}\label{fig001}
\begin{center}
\includegraphics[height=.35\textheight]{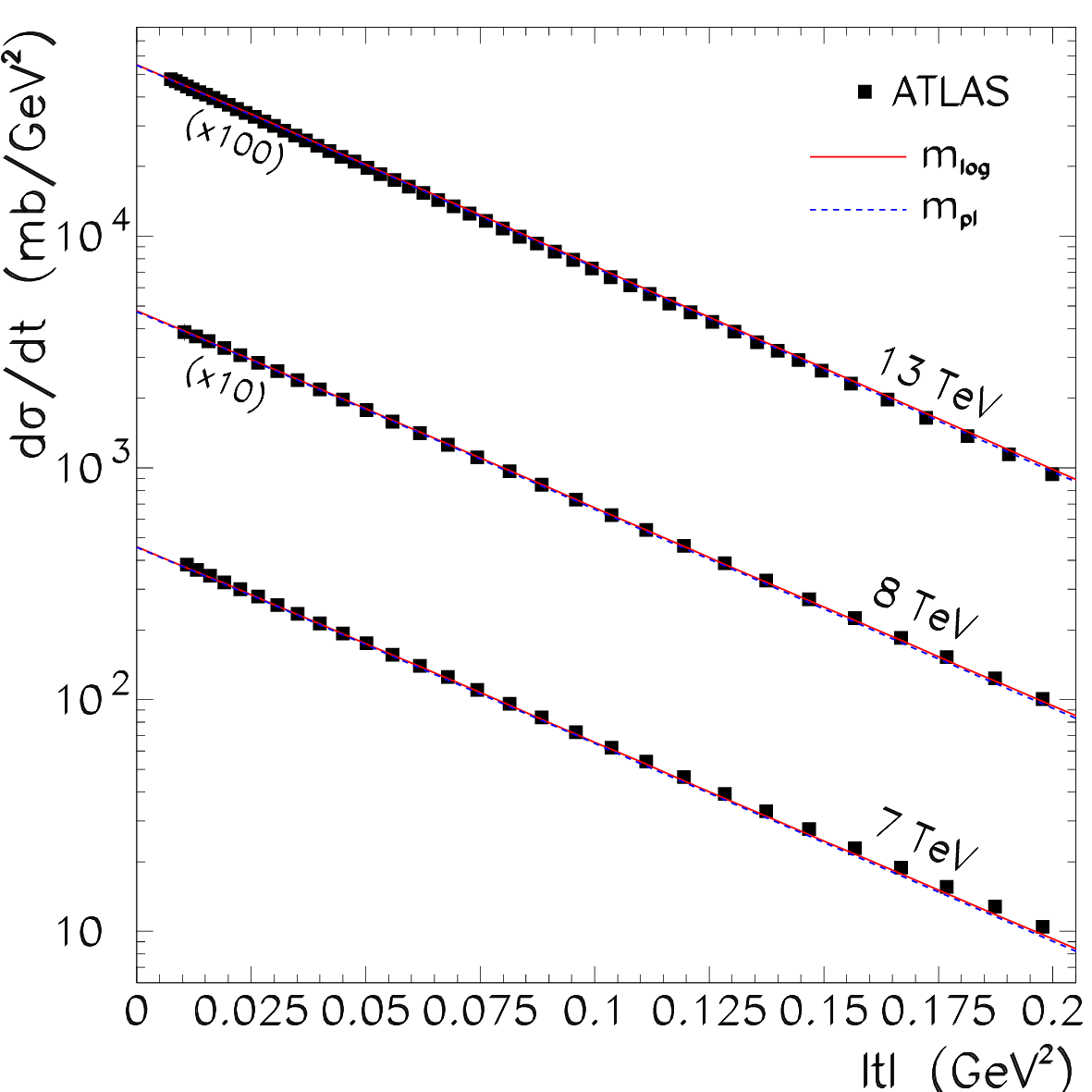}
\caption{LN Pomeron model description of the $pp$ elastic differential cross section data from ATLAS (Ensemble A). The solid and dashed lines show the results obtained using $m_{log}(q^{2})$ and $m_{pl}(q^{2})$, respectively.}
\end{center}
\end{figure}

\begin{figure}\label{fig002}
\begin{center}
\includegraphics[height=.35\textheight]{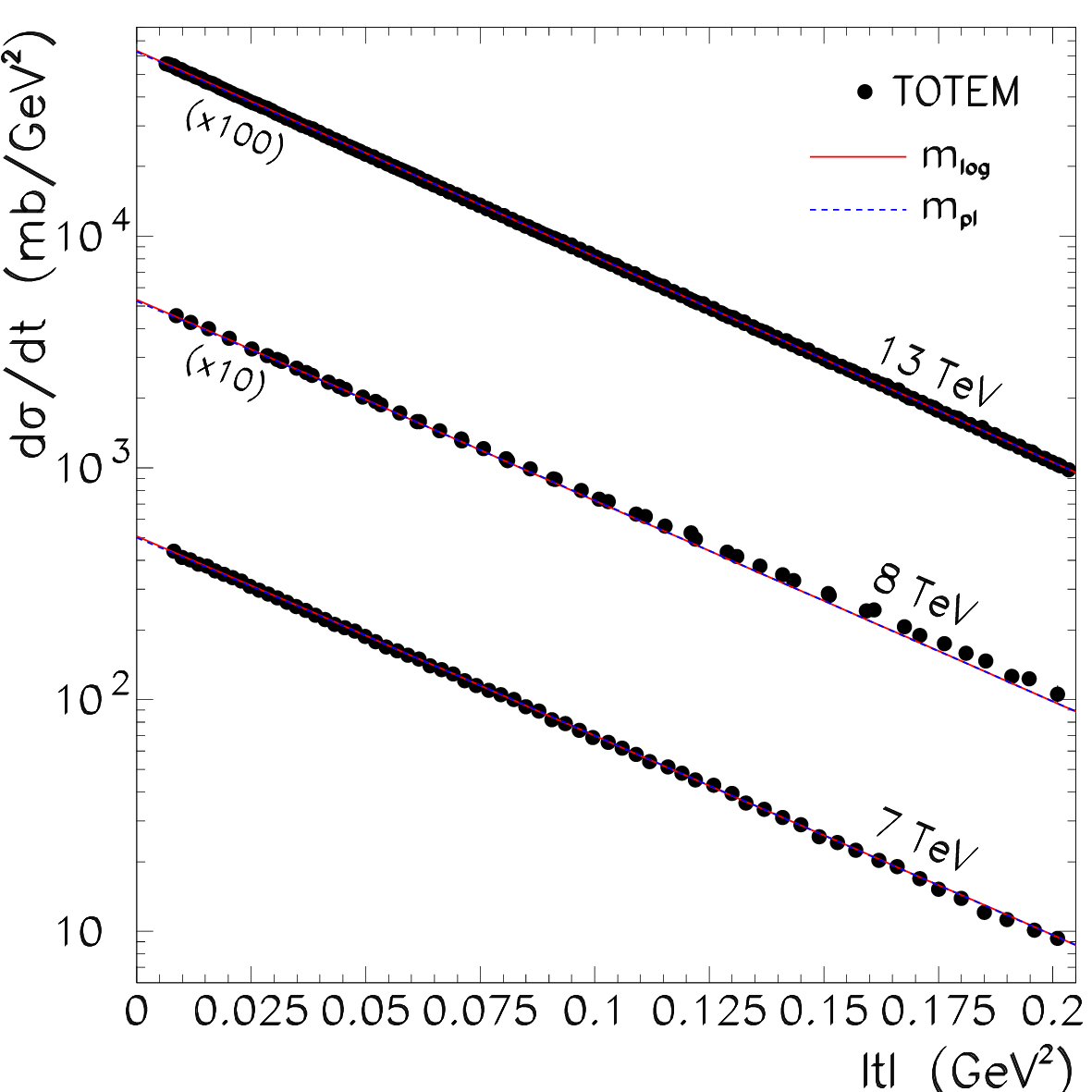}
\caption{LN Pomeron model description of the $pp$ elastic differential cross section data from TOTEM (Ensemble T). The solid and dashed lines show the results obtained using $m_{log}(q^{2})$ and $m_{pl}(q^{2})$, respectively.}
\end{center}
\end{figure}

\begin{figure}\label{fig003}
\begin{center}
\includegraphics[height=.35\textheight]{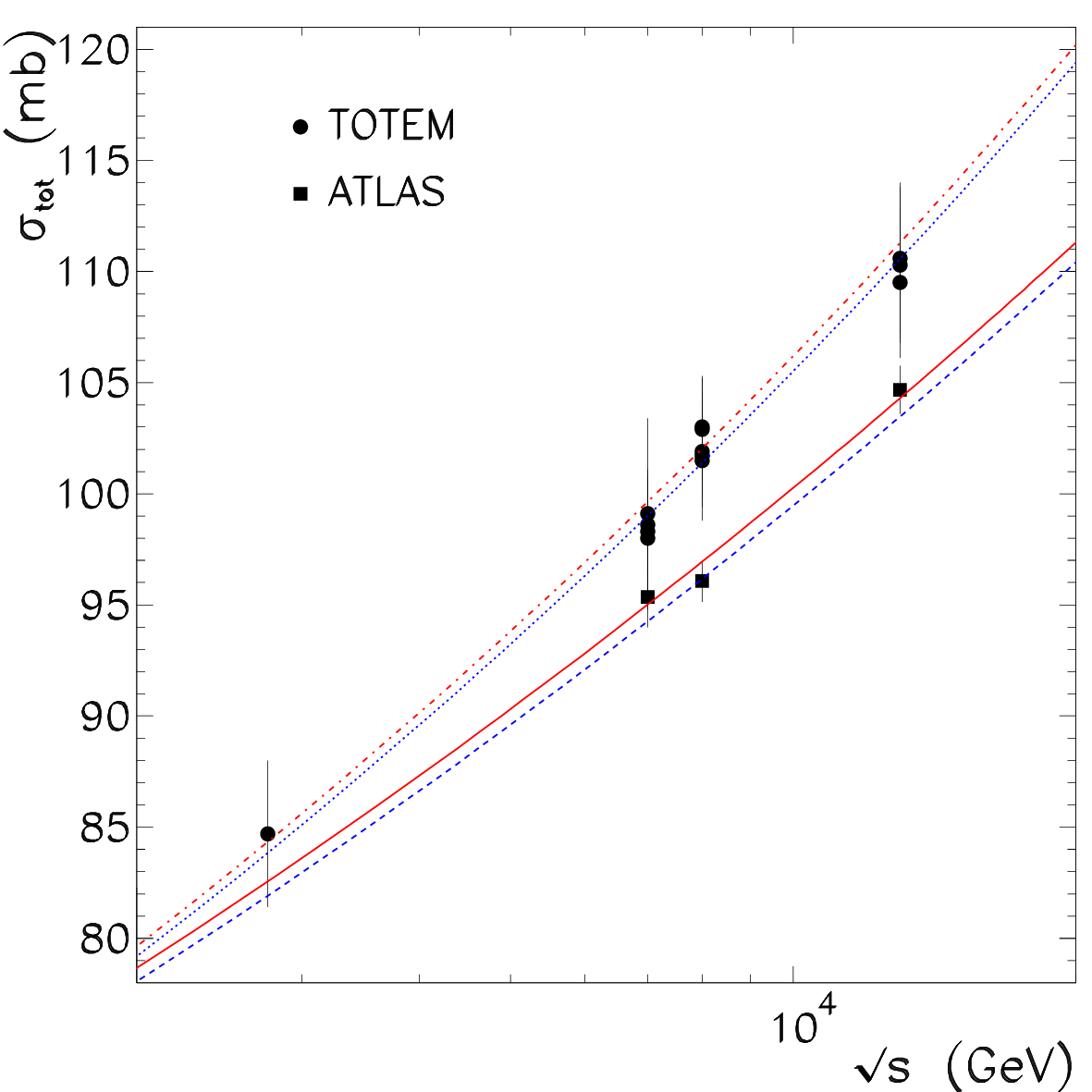}
\caption{LN Pomeron model prediction for the $pp$ total cross section. The solid, dashed, dash-dotted, and dotted lines are the predictions obtained from the fit to Ensemble A using $m_{log}(q^{2})$, Ensemble A using $m_{pl}(q^{2})$, Ensemble T using $m_{log}(q^{2})$, and Ensemble T using $m_{pl}(q^{2})$, respectively.}
\end{center}
\end{figure}

The energy dependence of the total and differential cross sections is driven by the parameter $\epsilon$, and we notice that for each given ensemble its value is not sensitive to the type of dynamical mass used in the fit: for the case of Ensemble A (Ensemble T), similar values of $\epsilon$, namely $\epsilon=0.075$ ($\epsilon = 0.089$), are obtained for both power-law- and logarithmic-type masses. As already advanced in the previous sections, the discrepancy between the values of $\epsilon$ obtained from distinct ensembles leads to different scenarios for the growth of the total cross section $\sigma_{tot}(s)$. Specifically, the model predictions for $\sigma_{tot}(s)$ at $\sqrt{s} = 13$ TeV in the case of Ensemble A using $m_{log}(q^{2})$, Ensemble A using $m_{pl}(q^{2})$, Ensemble T using $m_{log}(q^{2})$, and Ensemble T using $m_{pl}(q^{2})$, are approximately equal to 104.3 mb, 103.5 mb, 111.3 mb, and 110.9 mb, respectively. The curves of $\sigma_{tot}(s)$ corresponding to these four cases are shown in Figure 3.

We illustrate the behavior of the dynamical masses $m_{i}(q^{2})$ (Figure 4), the QCD effective charges $\bar{\alpha}_{i}$ (Figure 5), and the product $\bar{\alpha}_{i}(q^2)D(q^2)$ (Figure 6) in order to get a feeling for the sensitivity of the results on these quantities. Figures 4, 5, and 6 have the same legend as Figure 3.

\begin{figure}\label{fig004}
\begin{center}
\includegraphics[height=.35\textheight]{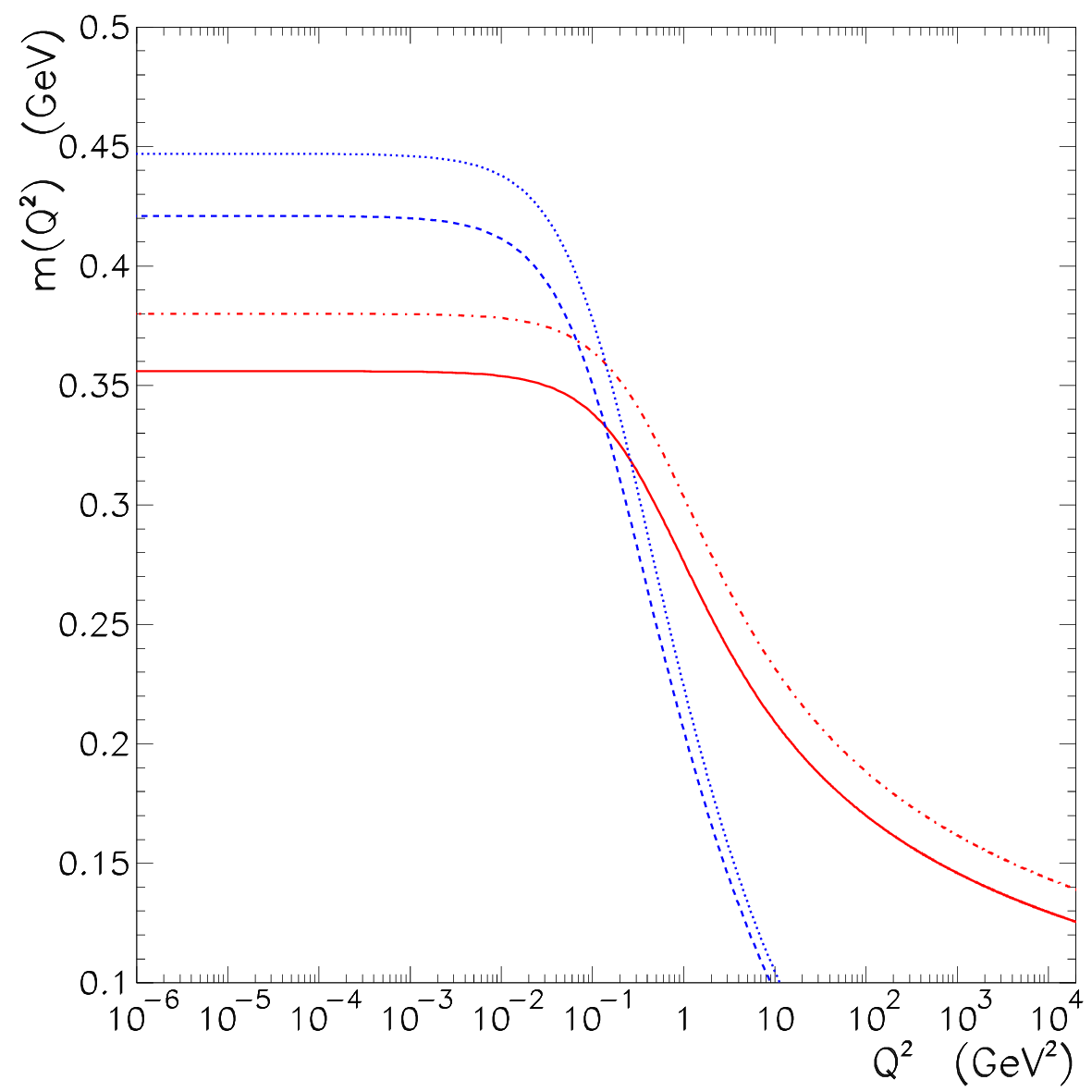}
\caption{The behavior of the dynamical masses. The solid, dashed, dash-dotted, and dotted lines are the masses observed using the parameters obtained from the fit to Ensemble A using $m_{log}(q^{2})$, Ensemble A using $m_{pl}(q^{2})$, Ensemble T using $m_{log}(q^{2})$, and Ensemble T using $m_{pl}(q^{2})$, respectively.}
\end{center}
\end{figure}

\begin{figure}\label{fig005}
\begin{center}
\includegraphics[height=.35\textheight]{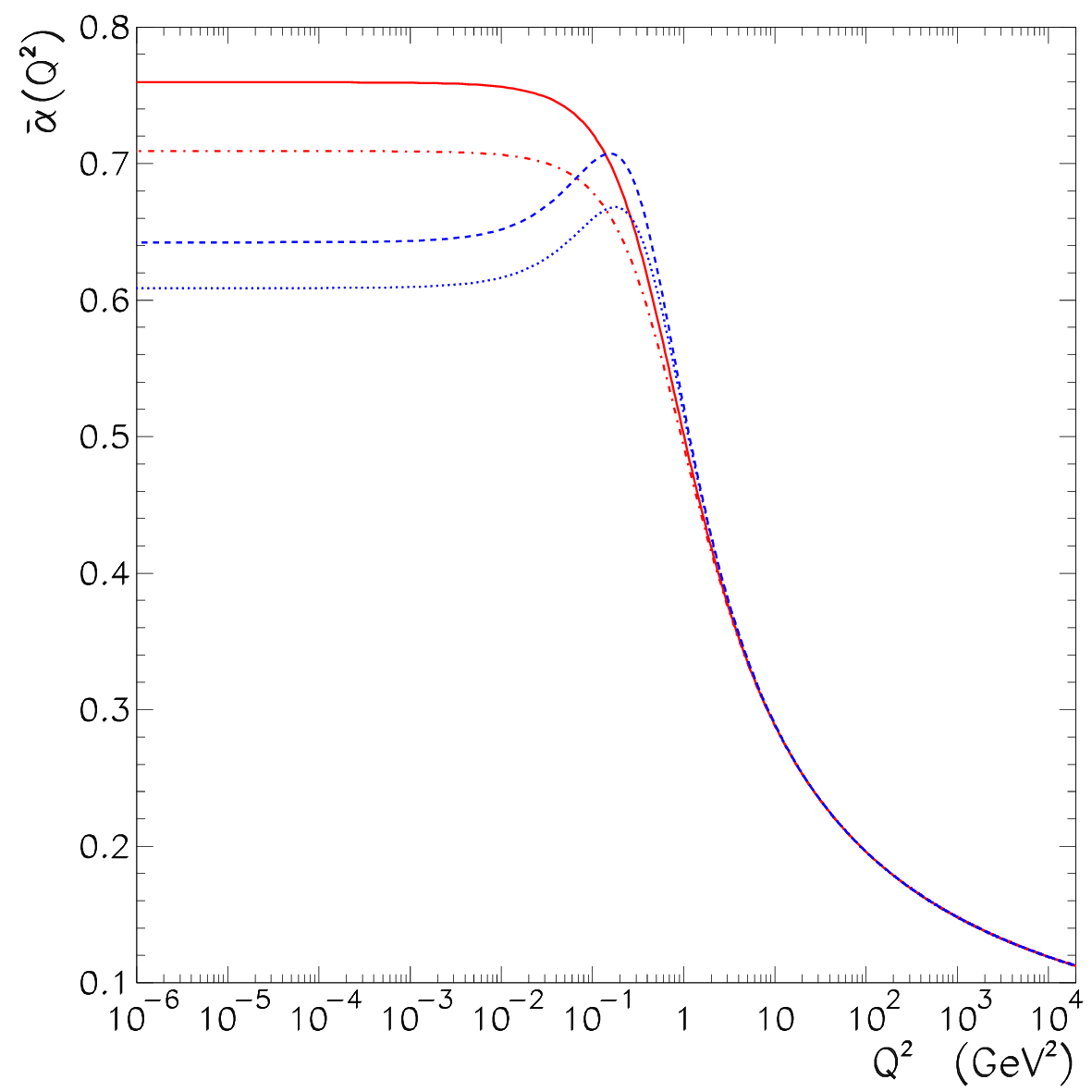}
\caption{The behavior of the QCD effective charges. The solid, dashed, dash-dotted, and dotted lines are the same as in Figure 4.}
\end{center}
\end{figure}

\begin{figure}\label{fig006}
\begin{center}
\includegraphics[height=.35\textheight]{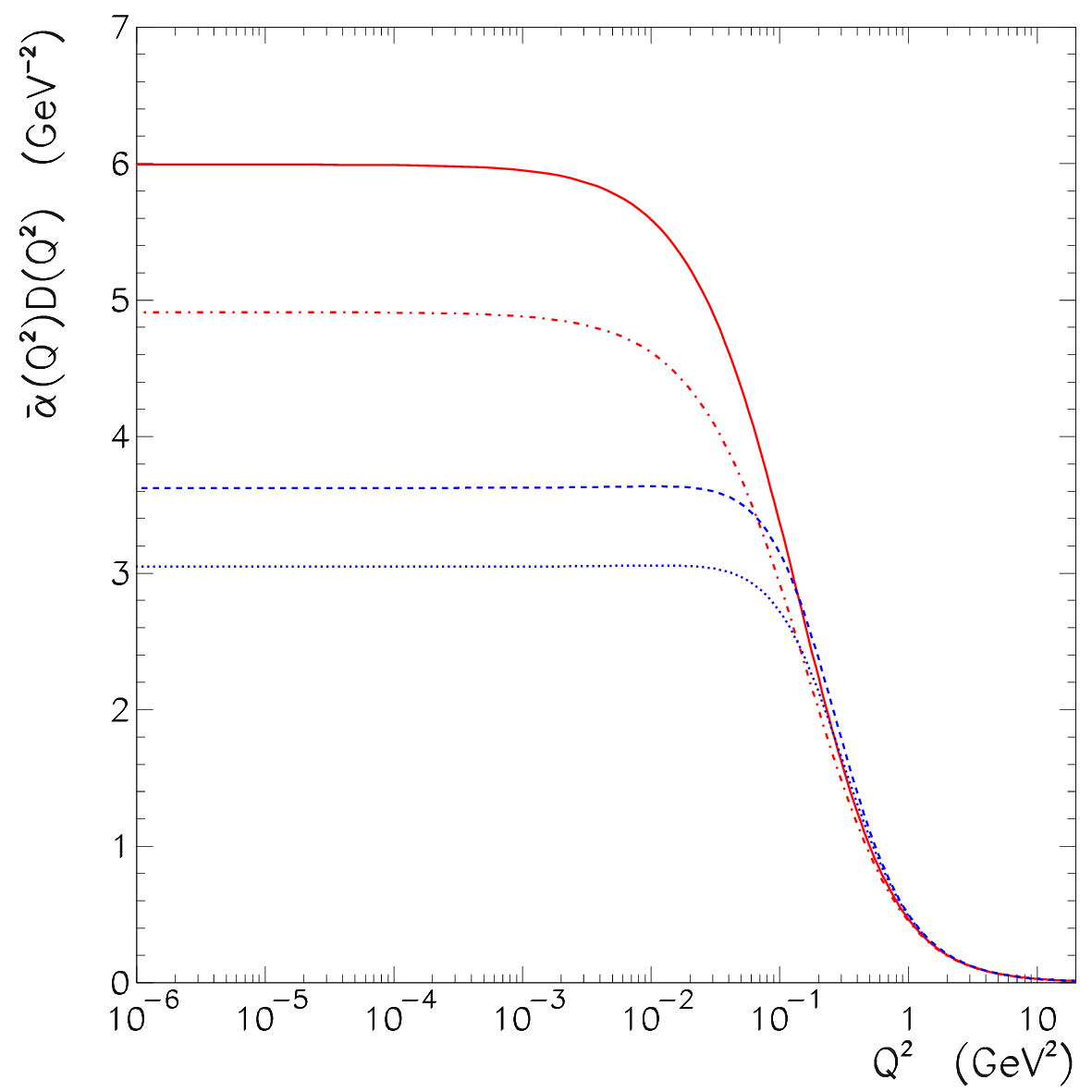}
\caption{The behavior of the product $\bar{\alpha}_{i}(q^2)D(q^2)$. The solid, dashed, dash-dotted, and dotted lines are the same as in Figure 4.}
\end{center}
\end{figure}

Interestingly enough, by considering the same type of dynamical mass, the change from Ensemble A to Ensemble B leads to an increase of $m_{g}$ of about 7\% and, by considering the same Ensemble, the change from the logarithmic to power-law mass leads to an increase of $m_{g}$ of some 18\%. The latter would normally be expected since power-law type masses decrease much faster than logarithmic ones, and this effect is exactly compensated by larger values of $m_{g}$.

Since we know the phenomenological values of the dynamical gluon mass, we are able to calculate the strength of the LN Pomeron coupling to quarks, given by the expression (\ref{eqln002}). For Ensemble A, in the case of logarithmic and power-law couplings, we have
\begin{eqnarray}
\beta_{0,\textnormal{\tiny ATLAS}} = 2.33^{+0.39}_{-0.30} \ \textnormal{GeV}^{-1},
\nonumber
\end{eqnarray}
\begin{eqnarray}
\beta_{0,\textnormal{\tiny ATLAS}} = 2.13^{+0.33}_{-0.25} \ \textnormal{GeV}^{-1} ,
\nonumber
\end{eqnarray}
respectively. On the other hand, for the Ensamble T, in the case of logarithmic and power-law couplings, we have
\begin{eqnarray}
\beta_{0,\textnormal{\tiny TOTEM}} = 2.04^{+0.28}_{-0.22} \ \textnormal{GeV}^{-1},
\nonumber
\end{eqnarray}
\begin{eqnarray}
\beta_{0,\textnormal{\tiny TOTEM}} = 1.91^{+0.22}_{-0.19} \ \textnormal{GeV}^{-1},
\nonumber
\end{eqnarray}
respectively. The uncertainty in these quantities has been estimated by varying the gluon mass $m_{g}$ within error while keeping all other model parameters constant. It is certainly obvious that this procedure does not determine the formal uncertainty of $\beta_{0}$.             
However, the values of $\beta_{0}$ are actually more sensitive to the gluon mass $m_{g}$ than to variations of other parameters of the model. Thus, despite the simplicity of the procedure, it clearly provides a reasonable estimate of the uncertainty in $\beta_{0}$.
It is worth mentioning that the expressions (19)-(22) are the expressions obtained from fits of Schwinger-Dyson equations solutions. The systematic exploration of the QCD Green’s functions through continuous Schwinger function methods has afforded broad access to the dynamical mechanisms responsible for the nonperturbative properties of the theory. On the other hand, to the best of our knowledge, the most recent QCD-Lattice result for the Pomeron’s strength was obtained in the quenched approximation, which amounts to neglecting quark loops. Moreover, the QCD running coupling was neglected in the lattice calculation, adopting the approximation $g_{eff}(p)=g$.
In this way, we consider the QCD-lattice result using the formula $\int d^{2}p\, [g^{2}_{eff}(p^{2})D_{lat}(p^{2})]^{2}$ only as a helpful guide, relying more on the intervals $\beta_{0,\textnormal{\tiny ATLAS}}$ and $\beta_{0,\textnormal{\tiny TOTEM}}$ calculated via Schwinger-Dyson formalism.

In conclusion, we verified that a two-gluon exchange model gives a very good description of the $d\sigma/dt$ data at TeV energies, provided we demand the Reggeization of the elastic scattering amplitude as predicted by QCD, and make a suitable choice for the convolution of proton wave functions at $k=0$. More precisely, we have evaluated the relative plausibility of different cumulant expansions for the form factor and, using two types of QCD effective charges (couplings), we have described for the first time high-energy differential cross sections data, in the interval $0 < |t|\leq 0.2$ GeV$^{2}$, using an LN inspired model.

We plan to extend our analysis to $d\sigma /dt$ data with $|t| > 0.2$ GeV$^{2}$ since it is generally believed that at large $|t|$ values the Odderon can play an important role \cite{odderon}. In performing calculations in the dip region, it is necessary to obtain the real part of the scattering amplitude, $\textnormal{Re}\, {\cal A}(s,t)$. Thus, it is essential the development of appropriate dispersion-relation techniques.
Further study of the behavior of other functional forms of the form factor becomes interesting at this stage of the work. For example, in References \cite{Iachello1,Iachello2}, experimental data on the nucleon's spacelike and timelike form factors were analyzed in terms of a two-component model for the electromagnetic form factor. Since electromagnetic and hadronic form factors have similar structures (both even having zeros in the same region in the momentum-transfer space \cite{elechadr1}), the study of hadronic form factors inspired by the electromagnetic two-component form factor and other electromagnetic functional forms becomes a natural extension of this work.
We also are interested in testing the sensitivity of our results to coupling constants that goes to zero in the deep infrared as observed by lattice simulations \cite{duarte001}. In particular, we are interested in the Curci-Ferrari gluon propagator and the coupling constant obtained from that approach \cite{oliveira909}.

\section*{Acknowledgments}

This research was partially supported by the Agencia Nacional de Investigaci\'on e Innovaci\'on under the project ANII-FCE-166479, by the Coordena\c{c}\~ao de Aperfei\c{c}oamento de Pessoal de N\'{\i}vel Superior (CAPES), and by the Conselho Nacional de Desenvolvimento Cient\'{\i}fico e Tecnol\'ogico (CNPq) under Grants No. 303588/2018-7 and No. 307189/2021-0.

\end{document}